\documentclass[fleqn,10pt]{wlscirep}
\usepackage{amsmath, amsthm, amssymb}
\usepackage{mathtools}

\usepackage{dcolumn}
\usepackage[tight]{subfigure}
\usepackage{gensymb}
\usepackage{verbatim}
\usepackage{inputenc}
\usepackage{xcolor}

\title{High operating temperature in V-based superconducting quantum interference proximity transistors.}

\author[1,*]{Nadia Ligato}
\author[1,2]{Giampiero Marchegiani}
\author[1]{Pauli Virtanen}
\author[1]{Elia Strambini}
\author[1,+]{Francesco Giazotto}

\affil[1]{NEST Istituto Nanoscienze-CNR and Scuola Normale Superiore, I-56127 Pisa, Italy}
\affil[2]{Dipartimento di Fisica dell'Universit\`{a} di Pisa, Largo Pontecorvo 3, I-56127 Pisa, Italy}
\affil[*]{nadia.ligato@nano.cnr.it} 
\affil[+]{francesco.giazotto@sns.it}

\begin{abstract}
Here we report the fabrication and characterization of fully superconducting quantum interference proximity transistors (SQUIPTs) based on the implementation of vanadium (V) in the superconducting loop. At low temperature, the devices show high flux-to-voltage (up to 0.52$\ \textrm{mV}/\Phi_0$) and 
flux-to-current (above 12$\ \textrm{nA}/\Phi_0$) transfer functions, with the best estimated flux sensitivity $\sim$2.6$\ \mu\Phi_0/\sqrt\textrm{Hz}$ reached under fixed voltage bias, where $\Phi_0$ is the flux quantum. The interferometers operate up to $T_\textrm{bath}\simeq$ 2 $ \textrm{K}$, with an improvement of 70$\%$ of the maximal operating temperature with respect to early SQUIPTs design. The main features of the V-based SQUIPT are described within a simplified theoretical model. Our results 
open the way to the realization of SQUIPTs that take advantage of the use of higher-gap superconductors for ultra-sensitive nanoscale applications that operate at temperatures well above 1 K. 
\end{abstract}

\begin{document}
	
\flushbottom
\maketitle
% * <john.hammersley@gmail.com> 2015-02-09T12:07:31.197Z:
%
%  Click the title above to edit the author information and abstract
%
\thispagestyle{empty}

\section*{Introduction}
%One of the main subjects of modern superconductivity was, and continues to be, its implementation as pivotal element in technology. 
Currently, the possibility to control electrical\cite{sohn1997,Clarke2004squid} and thermal transport\cite{Giazotto2012heat,Giazotto2006transport} in hybrid superconducting systems has generated strong interest for nanoscale applications, including metrology\cite{Solinas2015,Pekola2007}, quantum information\cite{Clarke2008}, quantum optics\cite{Nori2011}, scanning microscopy\cite{Vasyukov2013}, thermal logic\cite{Fornieri2016,Paolucci2016} and radiation detection\cite{Govenius2016}. 

In this scenario, the superconducting quantum interference proximity transistor (SQUIPT)\cite{Giazotto2010,Meschke2014} represents a concept of interferometer which shows suppressed power dissipation and extremely low flux noise comparable to conventional superconducting quantum interference device (SQUID)\cite{Clarke2004squid,tinkham1996introduction}. A SQUIPT consists of a short metal wire (i.e., a weak-link) placed in good electric contact with two superconducting leads defining a loop and a metal probe tunnel-coupled to the nanowire. As a consequence of the wire/superconductor contacts, superconducting correlations are induced locally into the weak-link through the proximity effect\cite{DEGENNES1966,Buzdin2005,Kim2012}. This results in a strong modification of the density of states (DOS) in the wire, where a minigap is opened\cite{leSueur2008}. The key factor of the device is the possibility to control the wire DOS and thus the electron transport through the tunnel junction, by changing the superconducting phase difference $\varphi$ across the wire-superconductor boundaries through an applied magnetic field which gives rise to a flux $\Phi$ piercing the loop area.   

The transparency of the nanowire/superconductor contacts plays a key role in the device sensitivity, because the induced minigap in the wire DOS is highly sensitive to the interface transmissivity, and decreases as the contacts become more opaque\cite{Hammer2007}. In this sense, it is convenient to realize SQUIPTs where the nanowire and the loop are made of the same superconducting material due to the higher quality of the contacts interface as well as the simpler fabrication process. Recently, the features of fully superconducting Al-based SQUIPTs have been theoretically and experimentally investigated\cite{Virtanen2016,Ronzani2016}. 

So far, SQUIPT configurations show a wide use of Al as the superconducting material\cite{Giazotto2010,Ronzani2014,D'Ambrosio2015}. Its popularity is mainly due to the simple and extensive know-how of Al film deposition, and due to its high-quality native oxide which allows the realization of excellent tunnel barriers through room-temperature oxidation. However, the low value of the Al critical temperature ($ T_\textrm{c}=$ 1.2 K) is synonymous with low operation temperatures, and the use of superconducting metals with higher $T_\textrm{c}$ is greatly desired for technological applications. The use of elemental metals such as vanadium (V) and niobium (Nb) is technologically demanding but would enable the possibility to significantly extend the SQUIPT working temperature. Nb has a high $T_\textrm{c}= $ 9.2 K, but also high melting point that requires more complex nanofabrication processes\cite{Samaddar2013,Jabdaraghi2016}. Vanadium is a group-$V$ transition metal, such as Nb, with a bulk $ T_\textrm{c}= $ 5.4 K, but its lower melting point allows easier evaporation\cite{Giazotto2009VCuV,Ronzani2013V,Quaranta2011,Beltram}.

An essential requirement for an optimal phase bias of the SQUIPT device is that the kinetic inductance of the superconducting ring, $L_\textrm{kin}^\textrm R$, be negligible compared to that of the nanowire, $L_\textrm{kin}^\textrm {NW}$, i.e. $L_\textrm{kin}^\textrm {NW}/L_\textrm{kin}^\textrm {R}\gg 1$\cite{Taddei2011}. This condition makes using refractory metals as the ring material less favorable, due to the typically higher values of their resistivity (see Supplementary Information) \cite{Annunziata2010,Luomahaara2014,Zhao2016}.

Here we report the fabrication and characterization of V-based SQUIPTs realized with a V-Al bilayer ring. On the one hand the V implementation on top of an Al-SQUIPT ring allows us to extend the maximal operating temperature up to $T\sim\ 2$ K, granting a significant improvement of the operating temperature range with respect to early Al-based SQUIPTs. On the other hand the Al layer acts as a "shunt inductor" to ensure a low value of the $L_\textrm{kin}^\textrm R$ for an optimal phase bias of the device.
At low temperature our interferometers show good magnetic sensing performance, with a maximum flux-to-voltage transfer function of $\sim0.5\ \textrm{mV}/\Phi_0$ and % under fixed current bias, and 
a maximum flux-to-current transfer function of $\sim 12\ \textrm{nA}/\Phi_0$, where $\Phi_0\simeq 2.068\times 10^{-15}\ \textrm{Wb} $ is the flux quantum. The maximum flux sensitivity $\sim 2.6\ \mu\Phi_0/\sqrt\textrm{Hz}$ is obtained under optimal voltage bias. 

The Section Results is organized as follows. In the Subsection Interferometers design we briefly discuss the design and the fabrication of the device. The electric characterization at low temperature is presented in the Subsection Transport spectroscopy. The Subsection Magnetic sensing performance is devoted to the magnetometric behaviour at low temperature, with an evaluation of the transfer functions and the flus sensitivity. In the Subsection Impact of the bath temperature the temperature evolution of the interferometers features is discussed. 
 
\section*{Results}
\subsection*{Interferometers design}
\begin{figure}[tp]
	\begin{centering}
		\includegraphics[width=1\textwidth]{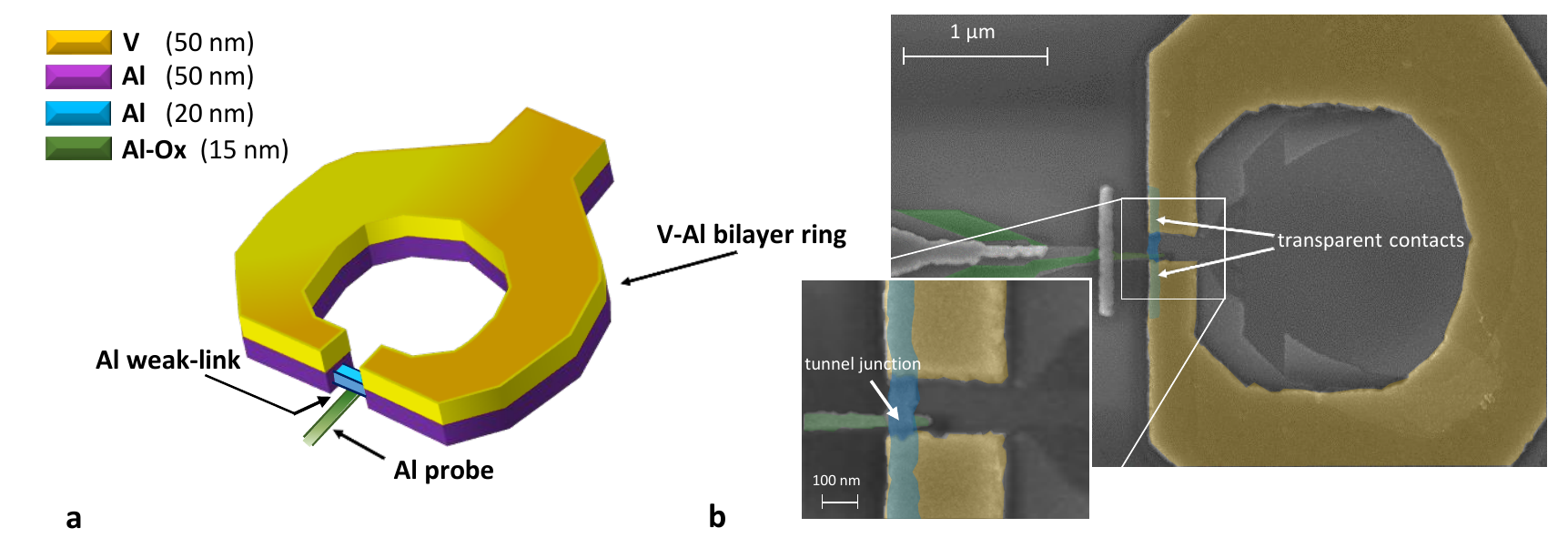}
		\caption{Design and scanning electron micrograph (SEM) of the V-based SQUIPT. (a) Sketch of the V-based SQUIPT. An Al nanowire is embedded into a V-Al ring and an Al probe is tunnel-coupled to the middle of the wire. (b) False-color SEM of sample-A with an enlarged view centered on the junction region. In the SEM image the passive metal replicas deriving from the three-angle shadow-mask evaporation are also visible.}
		\label{Fig1}
	\end{centering}
\end{figure} 
\begin{table}[b]
	\caption{Parameters of different SQUIPT samples. $L$ is the interelectrode spacing, $w_\textrm{NW}$ denotes the width of the superconducting nanowire, while $w_\textrm{pr}$ is the width of the Al probe. The tunnel resistance $R_\textrm T$, the maximum absolute values of the flux-to-current ($|dI/d\Phi|_\textrm{Max}$) and flux-to-voltage ($|dV/d\Phi|_\textrm{Max}$) transfer functions are measured at $T_\textrm{bath}=25\ \textrm{mK}$.}
	\begin{tabular}{c c c c c c c }
		\hline
		\hline
		\noalign{\smallskip}
		\vspace{0.1cm}
		& $L$ & $w_\textrm{NW}$ & $w_\textrm{pr}$ & $R_\textrm T$ & $|dI/d\Phi|_\textrm{Max}$ & $|dV/d\Phi|_\textrm{Max}$\\ 
		\vspace{0.1cm}
		Sample & (nm) & (nm) & (nm) & (k$\Omega$) & (nA$/\Phi_0$) & (mV$/\Phi_0$)\\ \hline
		A &150 & 60&30  & 56   & 12.0 & 0.52\\  
		B &140 & 65&30  & 61   & 10.5 & 0.49 \\ 
		C &155 & 50&40 & 65   & 8.3 & 0.48\\
		D &150 & 45&35 & 36   & 5.1 & 0.19\\ \hline\hline
		\vspace{0.1cm}
		\vspace{0.2cm}
	\end{tabular}
\end{table}   
The SQUIPT concept design [see Figure \ref{Fig1}(a)] is based on an Al nanowire embedded in a thick V-Al bilayer ring. Furthermore, an Al probe electrode is tunnel-coupled to the Al wire. The V layer deposited on top of the Al ring allows to increase the size of the minigap induced in the nanowire DOS, without compromising the $\textrm{Al}/\textrm{Al}\textrm{O}_\textrm x/\textrm{Al}$ junction quality. The Al layer was first deposited to insure the quality of the interface between the wire and the ring. 
The loop geometry of the superconducting electrode makes it possible to change the phase difference $\varphi$ across the superconducting wire by applying an external magnetic field, due to the flux quantization. The choice of a thick layer for the superconducting ring is a necessary condition: i) to reduce the inverse proximity effect of the Al wire on the bilayer ring and ii) to decrease its normal-state resistance, and thus its kinetic inductance, thereby allowing a good phase biasing of the weak-link. 

Interferometers are realized by electron-beam lithography (EBL) combined with three-angle shadow-mask evaporation (see Methods).
Figure 1(b) shows a false-color scanning electron micrograph (SEM) of a typical V-based SQUIPT with a magnification of the weak-link zone. A crucial step in the processing is the vanadium deposition. Electron-beam evaporation of a refractory superconductor material such as V requires some special considerations. If no precautions are taken, the heating of the substrate damages the resist layer with the consequent metal-pattern deterioration. In this regard, Table I lists the characteristic parameters for all samples and demonstrates the excellent reproducibility achieved as a consequence of the fabrication process optimization. 
 
\subsection*{Transport spectroscopy}

The SQUIPT operation relies on the magnetic-flux control of the weak-link DOS. For ideal wire/ring interfaces, the minigap $\epsilon_\textrm G$ in the middle of the wire in the short junction limit, i.e., when the interelectrode spacing $L$ is shorter than the diffusive coherence length $L\leq \sqrt{\hbar D/\Delta_\textrm R}$, is $\epsilon_\textrm G=\Delta_\textrm R |\cos(\varphi/2)|$. Here, $\hbar$ is the reduced Planck constant, $D$ is the diffusion coefficient of the nanowire, and $\Delta_\textrm R$ is the energy gap of the ring.
In the limit of negligible geometric and kinetic inductance of the ring compared to the weak-link kinetic inductance, the fluxoid quantization imposes $\varphi=2\pi\Phi/\Phi_0$ where $\Phi$ is the external magnetic flux piercing the loop.% and $\Phi_0=2.068\times 10^{-15}$ Wb is the flux quantum. 
\begin{figure}[t]
	\begin{centering}
		\includegraphics[width=1\textwidth]{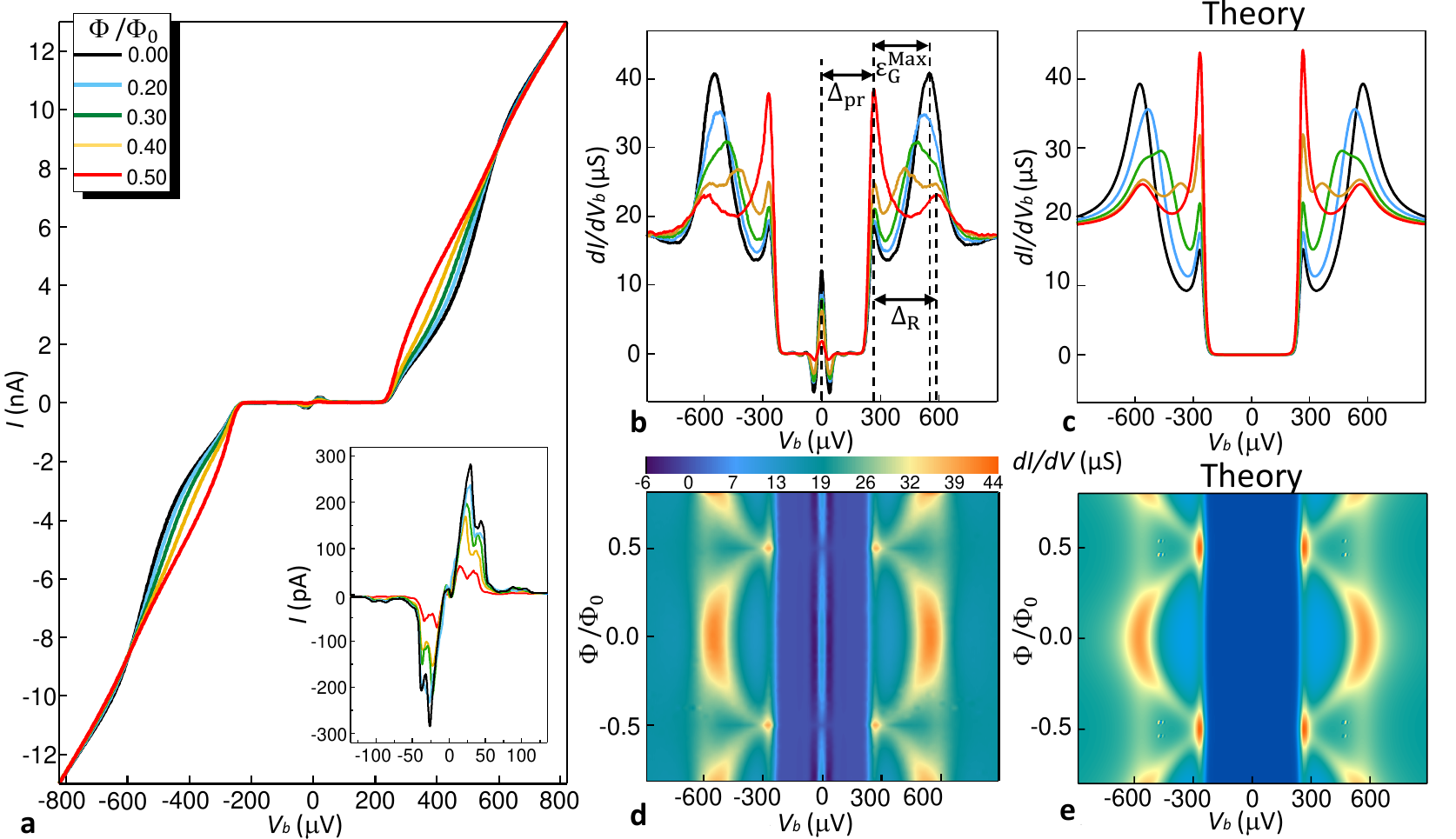}
		\caption{Interferometer characterization at $T_\textrm{bath}=25\ \textrm{mK}$ (sample-A). (a) Current-voltage $I(V_\textrm b$) characteristics measured for some values of the flux $\Phi$ generated by the external magnetic field. Inset: enlargement around zero bias of the $I(V_\textrm b)$ characteristics. The peak around zero bias, with maximum magnitude $I\simeq$ 280 pA, is the Josephson current flowing through the superconducting probe junction. (b) Measured and (c) theoretical differential conductance as function of voltage bias for $\Phi$ values as in (a). (d) Experimental and (e) theoretical color plot of the differential conductance $dI/dV$ versus voltage and magnetic flux.}
		\label{Fig2}
	\end{centering}
\end{figure}  
As a result, the electric transport through the leads is $\Phi_0$-periodic with the flux of the applied magnetic field. Thus the SQUIPT acts as an interferometer. All the measurements are performed in a $^3$He/$^4$He dilution refrigerator. The evolution of the electrical transport through the devices with the magnetic field is periodic with a period of $B_0 = 3.3$ G. The corresponding area $A_\textrm{eff}$ for magnetic field penetration is $A_\textrm{eff} = \Phi_0/B_0 \simeq6\ \mu\textrm m^2$, consistent with the size of the devices.

The characterization of the device (sample-A) at base temperature $T_\textrm{bath} =$ 25 mK is displayed in Fig.\ref{Fig2}.
Figure \ref{Fig2}(a) shows the $I(V_\textrm b)$ characteristics of the device measured for different values of the applied magnetic flux.
Here we can identify four regions: i) $|V_\textrm b|\leq 250\ \mu\textrm V$: the current is strongly suppressed and the phase modulation is negligible. ii) $250\ \mu \textrm V\leq |V_\textrm b|\leq 600\ \mu \textrm V$: the current increases significantly and a modulation with respect to the applied magnetic field is clearly visible.
iii) $|V_\textrm b|\geq 600\ \mu \textrm V$: a crossing point of the current-voltage characteristics and a small modulation is still visible. iv) at higher voltage the curves approach the ohmic behaviour. 
Moreover, since the probe is superconducting, the tunnel junction supports a supercurrent, due to the Josephson effect. This is shown in the inset of Figure \ref{Fig2}(a), where a magnification at low voltage is displayed. The supercurrent is $\sim 280$ pA at $\Phi=0$ where the minigap is maximum, while its value is reduced down to $\sim 60$ pA at $\Phi/\Phi_0=0.5$, showing a $\sim 80\%$ suppression with respect to the zero-flux value of the supercurrent. This behaviour gives an additional demonstration of the weak-link strong modulation of the minigap\cite{Ronzani2014}.

The magnetic field modulation of the nanowire DOS it is better visualized by considering the evolution of the differential conductance in the flux, as displayed in Figure \ref{Fig2}(b). 
The curves are obtained through numerical differentiation of the current-voltage characteristic shown in panel (a). Following to the I-V characteristics, the conductance is strongly suppressed for $|V_\textrm b|\leq 250\ \mu \textrm V$, except for the structures related to the Josephson effect. At $\sim 250\ \mu \textrm V$ an abrupt increase in the current [see Figure 2a] results in a conductance peak, whose intensity is enhanced by the applied magnetic flux. At higher absolute voltage values the peak evolution is more complex. At zero flux additional conductance peaks are visible at $|V_\textrm b|=550\ \mu \textrm V$. By increasing the magnetic flux, these peaks move toward smaller absolute voltages and their intensities become smaller, revealing the presence of additional structures at $580\ \mu \textrm V$.  

We explain this behaviour as follows. For simplicity, we consider only the quasiparticle contribution to the electrical current. The current through the probe/weak-link tunnel junction is given by\cite{Barone1982}

\begin{equation}
{I\text=\frac{1}{e R_\textrm T} \int_{-\infty}^{+\infty}dE\frac{1}{w_\textrm {pr}}\int_{x_0-\frac{w_\textrm {pr}}{2}}^{x_0+\frac{w_\textrm {pr}}{2}}dxN_\textrm{NW}(E,\Phi,x) N_\textrm{pr}(E-eV)[f(E-eV)-f(E)]}
\label{eq:thermocurrent}
\end{equation}
where $-e$ is the electron charge, $R_\textrm T$ is the normal-state resistance of the junction, $E$ is the quasiparticle energy with respect to the chemical potential and $V$ is the voltage across the junction. Here $N_\textrm{pr}$ and $N_\textrm {NW}$ are the normalized DOS %(normalized with the DOS at Fermi level) 
functions of the probe and the nanowire, respectively.
Since $w_\textrm{pr}/L\sim 0.2$, we approximate $\frac{1}{w_\textrm {pr}}\int_{x_0-\frac{w_\textrm {pr}}{2}}^{x_0+\frac{w_\textrm {pr}}{2}}dx N_\textrm{NW}(E,\Phi,x)$ $\approx$ $N_\textrm{NW}(E,\Phi,x_0)$, in order to simplify the calculation (see Supplementary Information).
The system is assumed to be at thermal equilibrium at temperature  $T_\textrm{bath}$ , thus the states population is expressed by the Fermi-Dirac distribution $f(E)=(e^{E/k_\textrm b T_\textrm{bath}}+1)^{-1}$.
The superconducting probe DOS is $N_\textrm{pr}(E)=N_\textrm{BCS}(E,\Gamma_\textrm{pr},\Delta_\textrm{pr})$, where $N_\textrm{BCS}(E,\Gamma,\Delta)=\left|{\mathrm{Re}\left[\frac{E+i\Gamma}{\sqrt{(E+i\Gamma)^2-\Delta^2}}\right]}\right|$ is the BCS DOS, smeared by a finite Dynes parameter $\Gamma$\cite{Dynes1984}. 
The nanowire DOS is affected by the proximity effect, which is properly described by the  quasiclassical Usadel equations for diffusive systems\cite{Usadel1970,Belzig1999}. 
In the short junction limit the solution for the DOS can be obtained analytically \cite{Heikkila2002}%\cite{Virtanen2016,Taddei2011}
\begin{equation}
N_\textrm {NW}(E,\Phi,x_0) =\left|\textrm{Re} \left[\frac{E+i\Gamma_\textrm R}{\sqrt{(E+i\Gamma_\textrm R)^2-\Delta_\textrm R^2 \cos^2(\pi\Phi/\Phi_0)}}
\cosh\left(\frac{2 x_0}{L}\cosh^{-1}\sqrt{\frac{(E+i\Gamma_\textrm R)^2-\Delta_\textrm R^2
		\cos^2(\pi\Phi/\Phi_0)}{(E+i\Gamma_\textrm R)^2-\Delta_\textrm R^2}}\ \right) \right]\right|,
\end{equation}
where the ring is modeled as an effective BCS superconductor with pairing potential $\Delta_\textrm R$ and Dynes parameter $\Gamma_\textrm R$. This expression simplifies in the limit of a perfectly centered probe ($x_0=0$), namely  $N_\textrm {NW}(E,\Phi,0)=N_\textrm{BCS}(E,\Gamma_\textrm R,\epsilon_\textrm G(\Phi))$, where $\epsilon_\textrm G(\Phi)=\Delta_\textrm R |\cos(\pi\Phi/\Phi_0)|$ is the flux-dependent minigap induced in the nanowire DOS. Similar applies at $\Phi=0$, where the wire DOS is independent on the probing position $N_\textrm{NW}(E,0,x_0)=N_\textrm {BCS}(E,\Gamma_\textrm R,\Delta_\textrm R)$. Notably, even for $\Phi=0.5\ \Phi_0$, the DOS retains a non-trivial dependence on the energy $E$ if $x_0\neq 0$.
 
Despite its simplicity, the model captures the main features observed in the differential conductance curves, including the evolution of the various peaks. The parameters $\Delta_\textrm{pr},\Delta_\textrm{R},\Gamma_\textrm{pr},\Gamma_\textrm{R}, x_0, R_\textrm{T}$ of the model can be separately estimated thanks to the rich structure expressed by the experimental curves.
The normal state resistance of the tunnel junction is easily extracted from the $I(V_\textrm b)$ characteristic in the ohmic limit (where $I\sim V_\textrm b/R_\textrm T $) as $R_\textrm T\simeq\ 56\ \textrm k\Omega$.
In the region $|V_\textrm b|\leq \Delta_\textrm{pr}/e\sim 250\ \mu \textrm V$ the conductance is strongly suppressed due to the superconducting energy gap in the probe DOS, whereas at higher voltages the conductance is large. This feature allows us to estimate both the Al probe pairing potential $\Delta_\textrm{pr}\simeq 255\ \mu e \textrm V$ and the ring Dynes parameter $\Gamma_\textrm R\sim 0.35\ \Delta_\textrm R $. Consequently, the Al probe Dynes parameter $\Gamma_\textrm {pr}\sim 10^{-3}\ \Delta_\textrm {pr} $ is determined from the small subgap conductance $\sim\Gamma_\textrm{R}\Gamma_\textrm{pr}/\Delta_\textrm{R}\Delta_\textrm{pr} R_\textrm T$. The peaks at voltage $|V_\textrm b|=(\Delta_\textrm{pr}+\Delta_\textrm R)/e\simeq\ 580\ \mu \textrm V$ which are visible at $\Phi=0.5\ \Phi_0$ allows for an estimation of the bilayer effective pairing potential $\Delta_\textrm R\simeq 310\ \mu\textrm{eV}$. Furthermore, it reveals a decentering in the probe position, which is set to $x_0=0.25\ L$, and is consistent with the scanning electron micrograph displayed in the enlarged view of the weak-link of Fig. \ref{Fig1} b). Summarizing, the three peaks structure at increasing voltage reside approximately at $e|V_\textrm b|\simeq\Delta_\textrm{pr},\ e|V_\textrm b|\simeq\epsilon_\textrm G(\Phi) + \Delta_\textrm{pr},\ e|V_\textrm b|\simeq \Delta_\textrm{R} +\Delta_\textrm{pr}$, where $\epsilon_\textrm G(\Phi)$ is the flux-dependent minigap induced in the nanowire DOS.

The theoretical curves for the differential conductance obtained using the above parameters are shown in Figure \ref{Fig2}(c). An extended comparison is presented in Figures \ref{Fig2} (d)-(e), where the color plots of the experimental and theoretical differential conductance are displayed, respectively. Note that the maximum minigap value $\epsilon_\textrm G^\textrm{Max}$  is slightly smaller than the ring pairing potential $\Delta_\textrm{R}$, namely $\epsilon_\textrm G^\textrm{Max}\simeq 300\ \mu e\textrm V$. This can be related to nonidealities in the  interface between the ring/weak-link contacts.

Two facts deserve discussion. First, it is difficult to give a precise estimate of the suppression of the minigap in the nanowire DOS at $0.5\ \Phi_0$ due to the presence of the probe pairing potential, which masks any possible small contribution around $|V_\textrm b|=\Delta_\textrm{pr}/e\simeq 255\ \mu e\textrm V$. Anyway, the strong flux-modulation of the tunnel probe supercurrent and the good agreement with the short-limit junction model confirm an almost full closure of the minigap. We also note that the incomplete suppression of the Josephson current at $\Phi=0.5\Phi_{0}$ could stem from the decentering of probe junction.
Despite this inconvenience, the choice of a superconducting probe is beneficial for improved magnetic sensor performance, as already shown in Al-based SQUIPTs \cite{D'Ambrosio2015,Ronzani2014}. Second, the large broadening parameter for the ring DOS $\Gamma_\textrm R$ must be regarded as an effective parameter in this simplified description. In particular the origin of such large conductance is likely related to the presence of vanadium. 
Similar high subgap conductances have been observed in several occasions in V-based tunnel junctions\cite{Seifarth1973,Noer1975,Dettmann1977,Gibson1989,Shimada2014}, as well as in Nb-based tunnel junctions\cite{Julin2016}.

\subsection*{Magnetic sensing performance}
\begin{figure}[tp]
	\begin{centering}
		\includegraphics[width=1\textwidth]{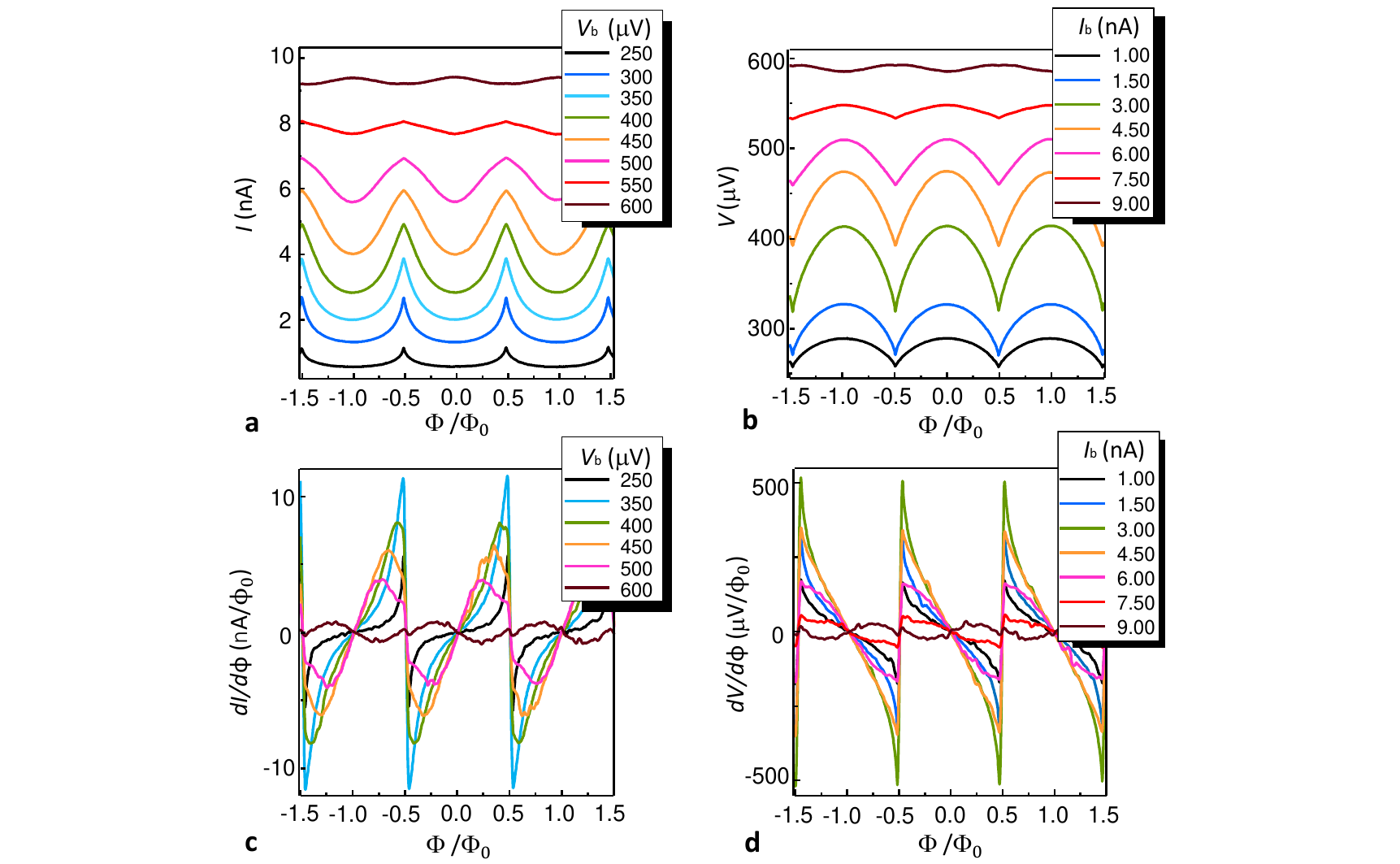}
		\caption{Interferometric behaviour of the V-based SQUIPT (sample-A) at $T_\textrm{bath}=25\ \textrm{mK}$. (a) Current modulation $I(\Phi )$ for different values of bias voltage $V_\textrm b$ applied to the tunnel junction. (b) Voltage modulation curves $V(\Phi)$ at different values of the bias current $I_\textrm b$ through the junction. (c) and (d) Flux-to-current $dI/d \Phi $ and flux-to-voltage $dV/d \Phi $ transfer functions, obtained differentiating $I( \Phi )$ and $V( \Phi )$, respectively.}
		\label{Fig3}
	\end{centering}
\end{figure}

 \begin{figure} [h]
	\begin{centering}
		\includegraphics[width=1\textwidth]{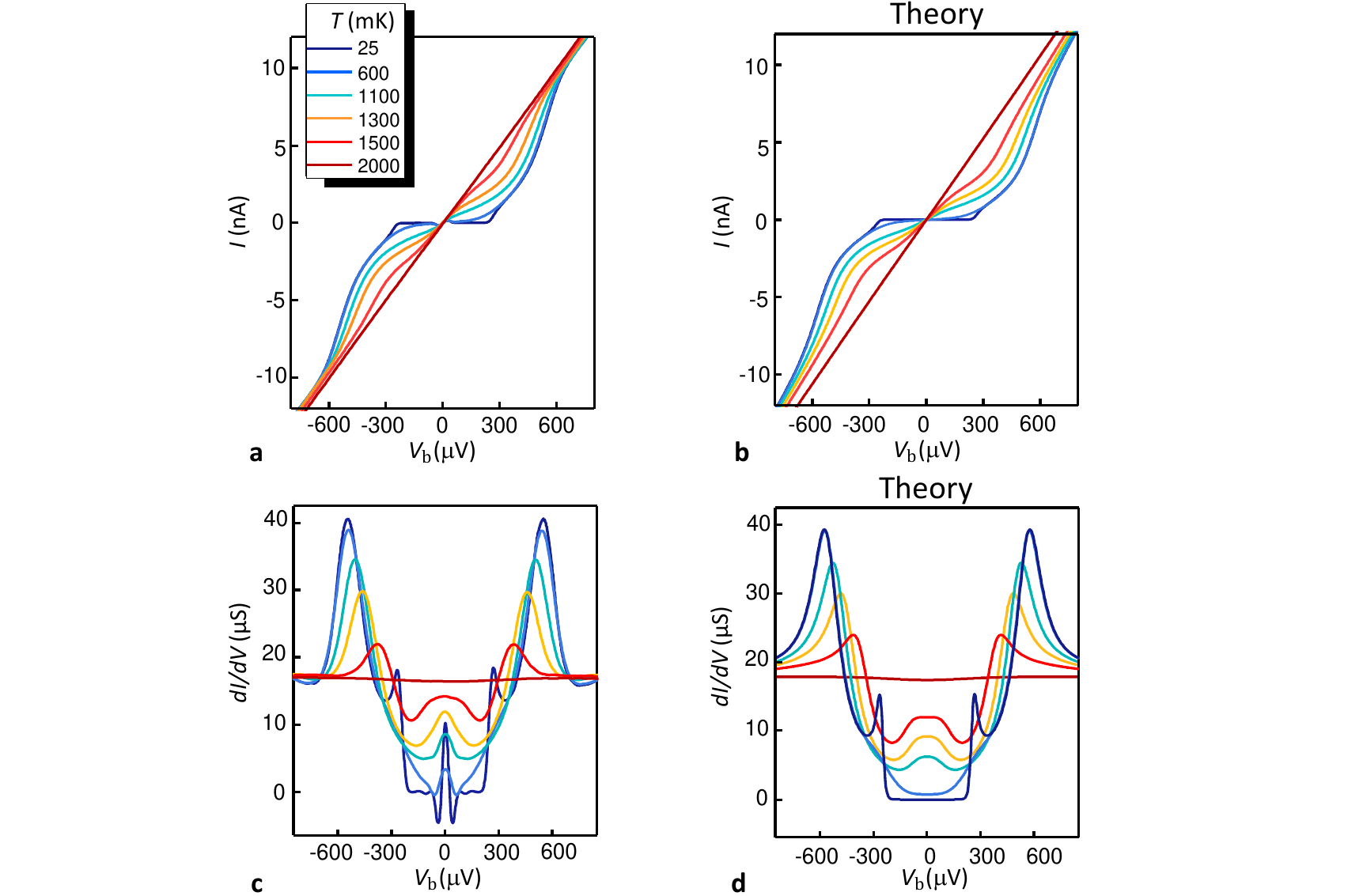}
		\caption{Temperature behaviour of the electrical trasport in the V-based SQUIPT. (a) Current-voltage $I(V_\textrm b$) characteristic curves measured at  $\Phi$=0 for several increasing temperatures of the V-SQUIPT (sample-A). (b) Theoretical prediction for the curves shown in (a), obtained through a numerical calculation based on the model presented in the text. (c) and (d) Measured and calculated differential conductance \textit{dI/dV} vs voltage bias for the same temperatures as in panel (a).}
		\label{Fig4}
	\end{centering}
\end{figure}
Here we investigate the interferometric behaviour of the sample-A SQUIPT at base temperature $T_\textrm{bath}=25\ \textrm{mK}$. For this purpose, we consider either the current modulation $I( \Phi )$ at fixed bias voltage $V_b$ and the voltage modulation $V( \Phi )$ at given input current $I_\textrm{b}$. The results are reported in Figure \ref{Fig3}. 
 
 Panel \ref{Fig3}(a) shows the current $I( \Phi )$ for several values of $V_\textrm b$  in the range from 250 $\mu$V to 600 $\mu$V, where the modulation is stronger. In accordance with the curves displayed in Figure \ref{Fig2}(a), the shape of $I(\Phi)$ and the size of the modulation strongly depend on the bias voltage $V_\textrm b$. The maximum current modulation $\delta I_\textrm{Max}$ in a period is approximately equal to $2\ \textrm{nA}$ around $V_\textrm b=400\ \mu$V. Note the change of concavity, i.e. the current decreases for increasing magnetic field in the range $[n \Phi_0,(n+1/2)\Phi_0$] ($n$ is an integer number), when $V_\textrm b$ exceeds the crossing point of the current-voltage characteristic [see Figure \ref{Fig2}(a)]. 
 
 In this configuration, the SQUIPT acts as a flux-to-current transducer. An important figure of merit for magnetic field sensing applications is the flux-to-current transfer function, namely $d I/d \Phi$. The curves obtained by numerical differentiation of the experimental data shown in Figure \ref{Fig3}(a) for six different values of bias voltage around the optimum working point are shown in \ref{Fig3}(c). The transfer function exhibits the maximum value of $|dI/d \Phi|_\textrm{Max}\cong12$ nA/$\Phi_{0}$ at $V_\textrm b=350 \ \mu \textrm{V}$. 
 A similar analysis is repeated for the current bias configuration. Figures \ref{Fig3}(b) and \ref{Fig3}(d) show the voltage modulation $V(\Phi )$ and the flux-to-voltage transfer function $d V/d \Phi$ for some values of the bias current $I_\textrm b$ in the range [1 nA,9 nA], respectively. In the half period [0,0.5$\Phi_0$] the voltage diminishes with the magnetic field due to the shrinking of the energy gap in the nanowire DOS, except for currents bigger than 8 nA, where the opposite occurs. The maximum voltage modulation $\delta V_\textrm{Max}$ and the maximum flux-to-voltage transfer function $|d V/d \Phi|_\textrm{Max}$ is obtained at 3.0 nA and reaches values as high as $\delta V_\textrm{Max}\cong100 \ \mu \textrm{V}$ and $|d V/d \Phi|_\textrm{Max}\cong520 \ \mu \textrm{V}/\Phi_{0}$, respectively.
 
 Another relevant figure of merit for a magnetometer is the noise-equivalent flux (NEF) or flux sensitivity ($\Phi_\textrm{NS}$), which gives the amount of noise per output bandwidth, and it is commonly expressed in units $\Phi_0/\sqrt\textrm{Hz}$. Thanks to the intermediate value of the tunnel junction resistance, the devices can efficiently operate either with voltage amplification under DC current bias or with current amplification under DC voltage bias. 

In the bias current configuration the flux sensitivity is expressed by 
$\Phi_\textrm{NS} =\sqrt{S_\textrm V}/|dV/d\Phi|$ where $S_\textrm V$ is the voltage noise spectral density. The intrinsic noise in the device is mainly given by the shot noise in the probe junction and it is expressed in the zero temperature limit by  $\sqrt {S_\textrm V}=R_\textrm d \sqrt{2eI}$  where $R_\textrm d=\partial V/\partial I$ is the differential resistance at the operating point.
The extrinsic noise is the input-referred noise power spectral density of the preamplifier used in this setup (NF Corporation model LI-75A, with $\sqrt{S_\textrm V}\sim 1.5\ \textrm{mV}/\sqrt\textrm{Hz}$). At the optimal bias point $I_\textrm b \sim 3\ \textrm{nA}$, $R_\textrm d\simeq 50\ \textrm k\Omega $ and the intrinsic and extrinsic noises give approximately the same contributions, reading $\Phi_\textrm{NS}\sim 3\ \mu \Phi_0/\sqrt{\textrm{Hz}}$ and $\Phi_\textrm{NS}\sim 2.9\ \mu \Phi_0/\sqrt{\textrm{Hz}}$, respectively. 

Improved performances are obtained in the voltage bias configuration, where the intrinsic noise is reduced to $\Phi_\textrm{NS} =\sqrt{2eI}/|dI/d\Phi|_\textrm{Max} \sim 2.6 \mu\Phi_0/\sqrt{\textrm{Hz}}$, where $I\sim 3\ \textrm{nA}$. In this configuration, the extrinsic contribution of the current preamplifier (DL Instruments model 1211, with current spectral density noise $\sqrt{S_I}=5\ \textrm{fA}/\sqrt{\textrm{Hz}}$) can be disregarded. In both configurations, the quantum limited noise $\Phi_{NS,q}= \sqrt{\hbar L_\textrm g}\sim 10 \textrm\ \textrm n\Phi_0/\sqrt{\textrm{Hz}}$ is negligible for typical ring geometric inductances $L_\textrm g\sim\ 5\ \textrm{pH}$.

\begin{figure}[tp]
	\begin{centering}
		\includegraphics[width=\textwidth]{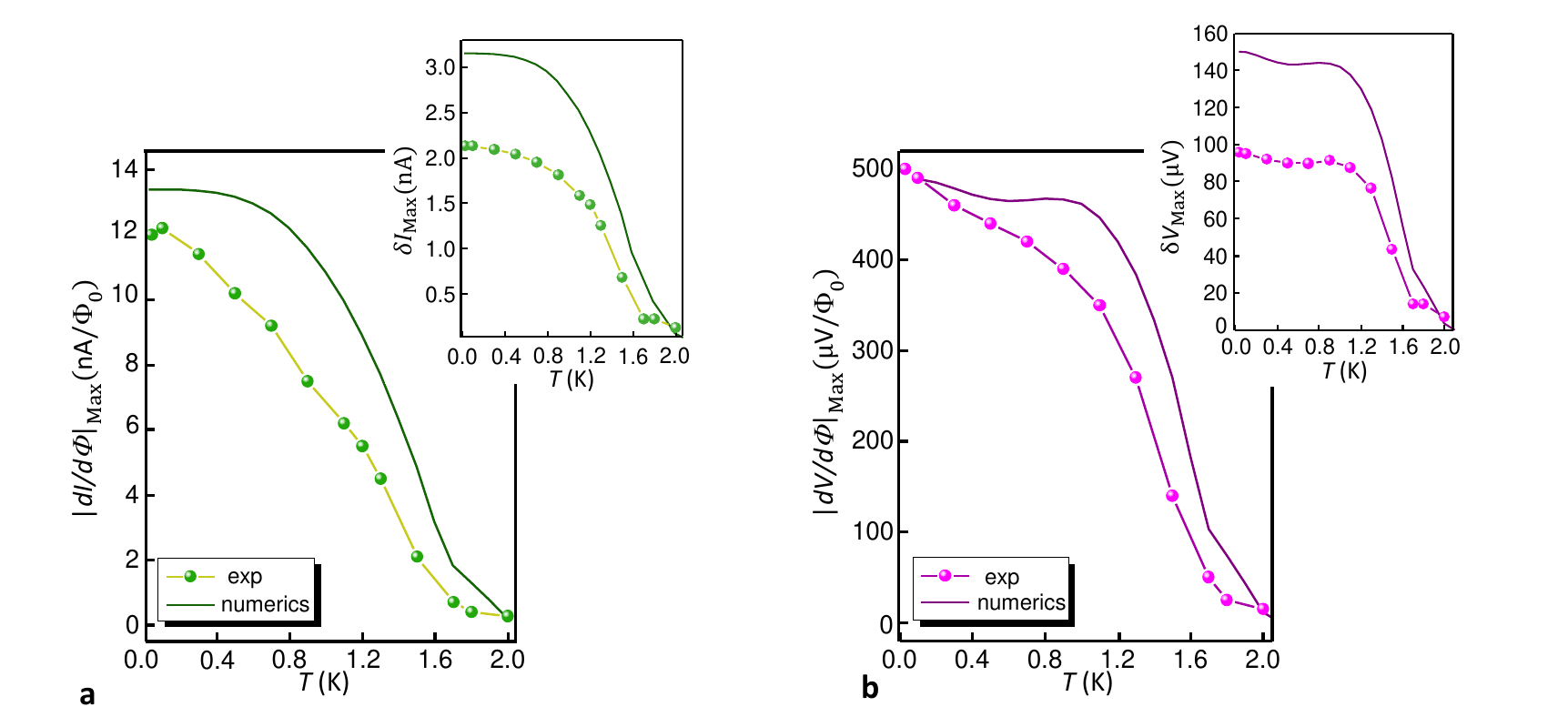}
		\caption{Temperature evolution of the magnetometer performance. (a) and (b) Temperature dependence of the maximum absolute value of the flux-to-current $(|dI/d\Phi|_\textrm{Max})$ and flux-to-voltage $(|dV/d\Phi|_\textrm{Max})$ transfer function, respectively. The insets show the maximum peak-to-peak current amplitude ($ \delta I_\textrm{Max}$) and the maximum peak-to-peak voltage amplitude ($\delta V_\textrm{Max}$), respectively. The lines connecting experimental data are guides to the eye. The theoretical predictions based on the simplified model are also plotted (solid lines)}.
		\label{Fig5}
	\end{centering}
\end{figure}
\subsection*{Impact of the bath temperature}
The role of the temperature $T$ is summarized in Figure \ref{Fig4}. Panel \ref{Fig4}(a) shows the $I(V_\textrm b)$ current-voltage characteristics at several bath temperatures for $\Phi =0$ (i.e., when the induced minigap is maximum). The corresponding theoretical curves are displayed in Figure \ref{Fig4}(b) for the parameters extracted from the base temperature characterization, and assuming a pure BCS dependence both for $\Delta_\textrm{pr}$ and $\Delta_\textrm R$. As expected, the current increases with the temperature due to the broadening of the Fermi distributions and the shrinking of the probe $\Delta_\textrm P$ and the ring $\Delta_\textrm R$ superconducting pairing potentials (hence the reduction of the minigap in the nanowire DOS $\epsilon_G$). Furthermore, the nonlinear behaviour of the $I(V_\textrm b)$ curves progressively decreases with the increase of the temperature, showing an almost linear characteristic around $2\ \textrm K$, when the superconducting features disappear. This value is consistent with the critical temperature extracted from the ring pairing potential $\Delta_\textrm R$, namely $T_\textrm{c,ring}= \Delta_\textrm R/(1.764 k_\textrm b)\simeq 2\ \textrm K$.

A deeper insight comes from the analysis of the differential conductances. Figure \ref{Fig4}(c) shows the curves obtained by numerical differentiation of the experimental data, while the corresponding theoretical curves are plotted in Figure \ref{Fig4}(d). Compared to the previously discussed features [see Figures \ref{Fig2}(b) and \ref{Fig2}(c)], the peak  structures have an easier identification, since the probe position $x$ has no effect on the nanowire density of states at $\Phi=0$, i.e., $N_\textrm {NW}(E,0,x)=N_\textrm{BCS}(E,\Gamma_\textrm R,\Delta_\textrm R)$.
At low temperature, i.e. for $T\ll T_{c,\textrm{probe}}=\Delta_\textrm{pr}/1.764 k_B\simeq 1.67\ \textrm K$ the Josephson supercurrent contribution appears as a low voltage peak. By increasing the temperature, this contribution becomes small due to the reduction of the probe and ring pairing potentials. In addition, it is masked by the quasiparticle contribution at low voltage, which becomes significant due to Fermi distribution broadening, thence it is spotted already at $1.1\ \textrm K$.

The peaks at $|V_\textrm b|=\Delta_\textrm{pr}/e\simeq 255\ \mu V$ are smoothed out by the thermal broadening already at $T=600$ mK, where only the peaks at higher absolute voltage are detectable.
As discussed before, the size of the minigap is related to the position of these peaks, which reside at $|V_\textrm b|=(\epsilon_\textrm G(\Phi)+ \Delta_\textrm{pr})/e$. As expected, their intensity decreases by increasing the temperature and they shift toward smaller absolute voltages, confirming the shrinking of the minigap, especially above 
$900\ \textrm{mK}\simeq 0.4\ T_{c,\textrm{ring}}$, according to the usual BCS dependence of the superconducting pairing potential. When the temperature reaches the $T_\textrm{c,ring}\simeq 2$ K, the differential conductance becomes almost flat. 

The temperature evolution of the interferometric performance of the device is shown in Figure \ref{Fig5}.
We already discussed how temperature increases the quasiparticle current, due to the thermal broadening and the shrinking of the probe and the ring gap. This smearing unavoidably influences the magnetic flux dependence of the current at fixed voltage and reduces the SQUIPT performance.
This aspect is shown in Figure \ref{Fig5}(a), where the maximum amplitude of the flux-to-current transfer function is displayed (points with line). This quantity decreases quite linearly with the temperature. Notably, the V-SQUIPT still exhibits a large sensitivity $|dI/d\Phi|_\textrm{Max}\simeq 2\ \textrm{nA}/\Phi_0$ at 1.5 K. This represents a relevant improvement with respect to the previous SQUIPT devices, where similar values were only possible below $1$ K. In the inset, the maximum current modulation at fixed voltage is plotted against temperature, showing a swing $\delta I_\textrm{Max}\simeq 700$ pA at 1.5 K. Similar considerations apply for the maximum flux-to-voltage $|dV/d\Phi|_\textrm{Max}$ and maximum voltage modulation $\delta V_\textrm{Max}$ at fixed current bias. The results are reported in Fig. \ref{Fig5}(b). These curves decrease slowly for temperatures $T\leq 0.4 T_\textrm{c,ring}\simeq 0.9$ K and then drop when the temperature approaches the bilayer critical temperature. As discussed before, the performances at 1.5 K are still remarkable, with a maximum transfer function $|dV/d\Phi|_\textrm{Max}\simeq 140\ \mu \textrm V/\Phi_0$ and a maximum swing $\delta V_\textrm{Max}\simeq 40\ \mu \textrm V$. In the plot the theoretical prediction, according to the simplified model used throughout the paper, are also reported (solid lines). The temperature evolution of the maximum flux-to-current (flux-to-voltage) transfer functions are similar to the experimental result, whereas significant deviations are observed in the current (voltage) swing. Our simplified model gives a somewhat less satisfactory fit for the current $I(\Phi)$ and voltage $V(\Phi)$ magnetic flux dependence than for the differential conductance (see Supplementary Information).

\begin{figure}[tp]
	\begin{centering}
		\includegraphics[width=\textwidth]{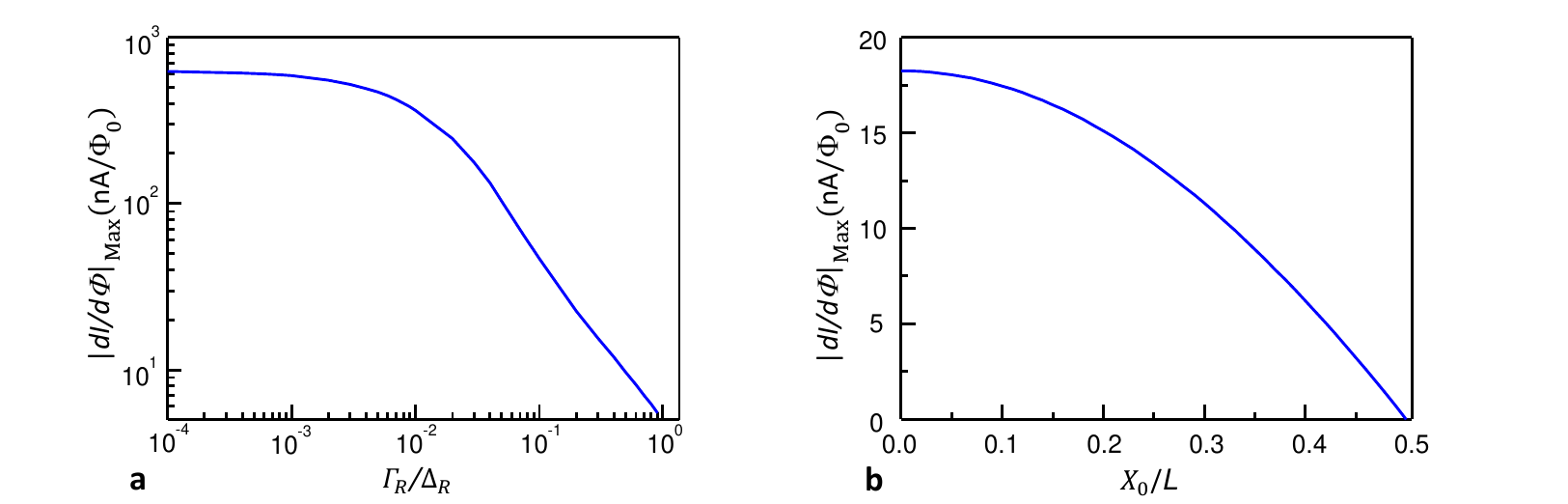}
		\caption{Theoretical prediction of the parameters impact on interferometers performance. (a) Maximum absolute value of the flux-to-current transfer function $(|dI/d\Phi|_\textrm{Max})$ versus normalized Dynes parameters of the ring. (b) Maximum absolute value of the flux-to-current transfer function $(|dI/d\Phi|_\textrm{Max})$ versus the probe position respect to the middle of the nanowire.}
		\label{Fig6}
	\end{centering}
\end{figure}
\section*{Discussion}
In summary, we have presented the fabrication and characterization of V-based SQUIPTs realized with a V-Al bilayer ring. Our quantum interference proximity transistors show good magnetometric performance with low noise-flux sensitivity down to $\sim2.6\ \mu\Phi_0/\sqrt\textrm{Hz}$ at base temperature $T_\textrm{bath}=25\ \textrm{mK}$. Previously, higher interferometric performances have been reported for Al-based SQUIPTs (magnetic flux resolution as low as $500\ \textrm n\Phi_0/\sqrt{\textrm{Hz}}$)\cite{Ronzani2014}. From the theoretical side, the V-based SQUIPT should guarantee better performance compared to the Al-based technology, since the predicted flux-to-voltage transfer function scales as $\Delta_\textrm R/e$\cite{Virtanen2016}. 
This is not observed due to the high subgap conductance of the superconducting vanadium, which limits the device performance. This is displayed in the first panel of Fig.\ref{Fig6}, where the theoretical maximum flux-to-current transfer function is plotted as a function of the ring Dynes parameter. A smaller role is played also by the decentering of the probe respect to the middle of the nanowire, as shown in Fig. \ref{Fig6}(b). The origin of the large subgap conductance of vanadium is still not understood and may be related to the evaporation process.
Despite this, the use of vanadium allows us to obtain unprecedented performances in terms of maximum operating temperature ($T\simeq 2.0\ \textrm K$), since the Al-based SQUIPTs typically work only up to $\sim$1 K.
Furthermore, our interferometers still exhibit high sensitivity ($|dI/d\Phi|_\textrm{Max}\simeq 2\ \textrm{nA}/\Phi_0$  and $|dV/d\Phi|_\textrm{Max}\simeq 140\ \mu\textrm{V}/\Phi_0$) at $1.5 $ K. The main features in our devices are well reproduced within a simplified theoretical model. 
The V-based SQUIPT configuration is a proof-of-concept showing the V-Al material combination is an suitable candidate for the realization  of high-performance magnetometers operating above $1$ K.

Finally, furthermore improvements will be possible with the adoption of superconducting materials with wider energy-gap, such as lead, niobium and niobium nitride. This will extend the magnetometer working operation at higher temperatures, allowing SQUIPT applications at temperatures accessible with the technology of $^4$He cryostats, in order to try to compete with the state of the art nanoSQUID\cite{Granata2016,Schmelz2017} (flux resolution $45\ \textrm n\Phi_0/\sqrt\textrm{Hz}$).

\section*{Methods}
\subsection*{Device fabrication details}
The devices were fabricated through single electron-beam lithography process followed by a three-angle shadow-mask evaporation of metals through a suspended resist mask. At first, an oxidized silicon wafer was covered with a suspended bilayer resist mask (1200-nm copolymer, 250-nm polymethyl methecrylate (PMMA)) by spin-coated process. Then the device structures have been patterned onto the substrate via electron-beam lithography. The EBL step is followed by development in 1:3 mixture of MIBK:IPA (methyl isobutyl ketone:isopropanol) for typically 1 min and 30 sec, followed by a rinse in pure IPA and drying with nitrogen. Then, the sample was processed in a ultra-high vacuum (UHV) evaporator (base pressure of $ 10^{-10}\ \textrm{Torr} $) for the metalization process. 15 nm of Al was deposited at 1.5\AA/s at an angle of $ \theta$=40\degree\ to form the superconducting electrode of the probe tunnel junction. Subsequently, the sample was exposed to 100 mTorr $ \textrm O_{2} $ for 5 min to realize the tunnel barrier. Next, the sample was tilted to an angle of  $ \theta $=20\degree\  for the evaporation of 20 nm of Al to form the superconducting nanowire. Subsequently, 50 nm of Al was deposited at $ \theta $=0\degree\ to realize the first layer of the bilayer ring. Finally, at the same angle 50 nm of V was evaporated at 3\AA/s in order to realize the upper layer of V/Al superconducting ring.
The magneto-electric measurements were performed in a $ ^{3} $He/$ ^{4} $He dilution refrigerator in a range temperature from 25 mK to 2 K using room-temperature preamplifiers. 

%\bibliography{references}

\section*{Acknowledgments}
The authors thank F. Paolucci, E. Enrico, and A. Fornieri for fruitful discussions. The MIUR-FIRB2013 – Project Coca (Grant No. RBFR1379UX) and the European Research Council under the European Union’s Seventh Framework Program (FP7/2007-2013)/ERC Grant agreement No. 615187- COMANCHE are acknowledged for partial financial support. The work of E.S. is funded by a Marie Curie Individual Fellowship (MSCA-IFEF-ST No. 660532-SuperMag).
\section*{Author contribution statement}
F.G. conceived the experiment. N.L. fabricated the samples. N.L., G.M. and E.S. performed the measurements. G.M. and P.V. developed the theoretical model. N.L and G.M analyzed the data. N.L. and G.M. wrote the manuscript. All authors reviewed the manuscript.
\section*{Additional Information}
The authors declare no competing financial interests.

\newpage
\setcounter{equation}{0}
\setcounter{figure}{0}
\setcounter{table}{0}
\setcounter{page}{6}
\makeatletter
\renewcommand{\theequation}{S\arabic{equation}}
\renewcommand{\thefigure}{S\arabic{figure}}

\section*{\bf Supplementary Information: High operating temperature in V-based superconducting quantum interference proximity transistors.}
\vspace{0.3cm}
\subsection*{Nadia Ligato, Giampiero Marchegiani, Pauli Virtanen, Elia Strambini, Francesco Giazotto}
\vspace{1cm}
\section*{Kinetic Inductance for the ring and the wire of the SQUIPT}

Within the Mattis-Bardeen theory\cite{tinkham2004introduction}, the kinetic inductance $L_\textrm {kin}$ of a superconducting strip with length $l$, width $w$ and thickness $t$ is given by $ L_\textrm {kin}=\hbar R/\pi\Delta$, where $\Delta$ is the superconducting order parameter and $R=\rho l/w t$ is the normal state resistance of the superconducting strip (here $\rho$ is the normal state resistivity). This expression can be used to estimate the kinetic inductance of the superconducting loop of the SQUIPT when the ring consist of a single superconductor.

For a comparison, we consider a sinusoidal current-phase relation for the superconducting weak-link, which in the short junction limit is valid when the temperature is not too small compared to the critical temperature\cite{Likharev1979}. 
Under this assumption, the minimal kinetic inductance at zero phase bias $\phi=0$ reads $ L_\textrm {kin}^\textrm{NW}\approx\hbar R_{NW}/\pi\Delta$,
where $R=\rho^\textrm{NW} l^\textrm{NW}/w^\textrm{NW} t^\textrm{NW}$ is the normal state resistance of the weak link.

The ratio between the kinetic inductance of the wire and the ring is
\begin{equation}
\frac{L_\textrm{kin}^\textrm{NW}}{L_\textrm{kin}^\textrm{R}}=\frac{\rho^\textrm{NW} l^\textrm{NW} t^\textrm{R} w^\textrm{R}}{\rho^\textrm{R} l^\textrm{R} t^\textrm{NW} w^\textrm{NW}}
\end{equation}
where the superscripts NW, R refer to the nanowire and ring, respectively.

The dimensions of the Al nanowire (device A) are $l^\textrm{NW}=150$ nm, $t^\textrm{NW}=20$ nm and $w^\textrm{NW}=60$ nm. We assume $\rho^{NW}\simeq 5\ \mu\Omega\ \textrm{cm}$, which is the typical resistivity for 25 nm Al layer at 4.2 K evaporated in past experiments, consistently with the values reported in the literature\cite{Nahum2000,Courtois2008,Beckmann2010,Peltonen2011}. If we consider a ring made of Al with dimensions $l^\textrm{NW}=6\ \mu$m, $t^\textrm{NW}=50$ nm and $w^\textrm{R}=1\ \mu$m  and same resistivity (although in general the resistivity drops by increasing the thickness of the layer) we obtain
$L_\textrm{kin}^\textrm{NW}/L_\textrm{kin}^\textrm{R}\sim 1.05$. 
\begin{figure}[h]
	\begin{centering}
		\includegraphics[width=\textwidth]{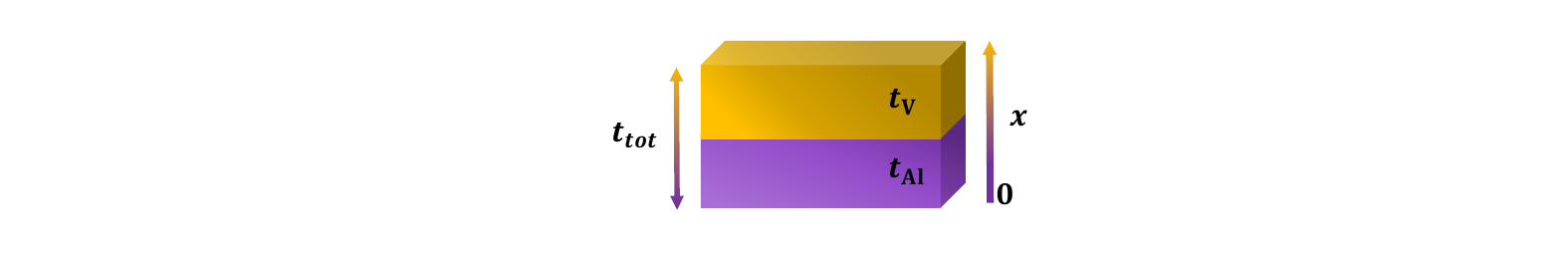}
		\caption{V-Al bilayer scheme. The strip has total thickness $t_\textrm{tot}=t_\textrm{V}+t_\textrm{Al}$, where $t_\textrm{V}$ and $t_\textrm{Al}$ are the thicknesses of vanadium and aluminium, respectively.}
		\label{FigS1}
	\end{centering}
\end{figure}
%On the contrary, 
The resistivity of the vanadium may vary quite strongly depending on evaporation conditions. Considering literature values \cite{Teplov1976,Nicolet1978,Kanoda1987,Gibson1989,Geers1997}, we estimate the V layer resistivity $\rho_\textrm{V}$ to range approximately from the same resistivity of the Al $5\ \mu\Omega\ \textrm{cm}$ to a value 5 times larger  $25\ \mu\Omega\ \textrm{cm}$. As a consequence, this would produce a potentially large deviation from the ideal condition $L_\textrm{kin}^\textrm{NW}/L_\textrm{kin}^\textrm{R}\gg 1$.
\begin{figure}[h]
	\begin{centering}
		\includegraphics[width=\textwidth]{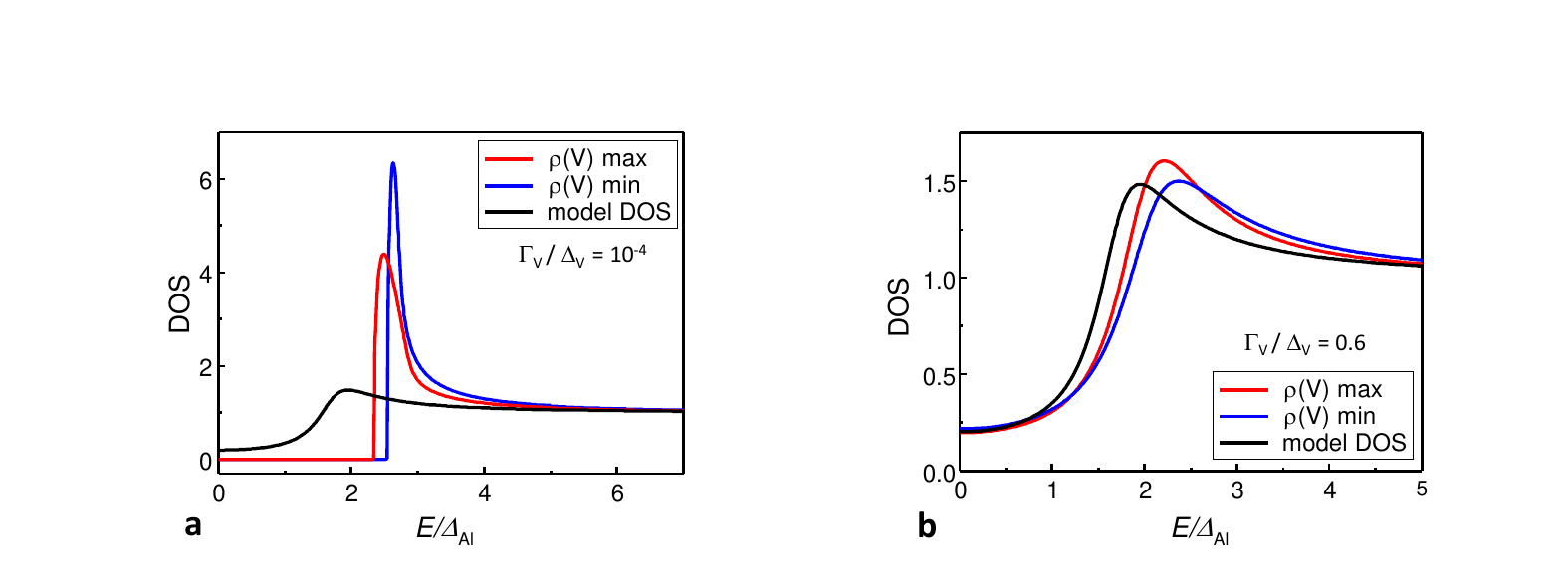}
		\caption{Density of states at the bottom of the Al layer for a 100 nm V-Al bilayer with thickness ratio 1:1. The curves for 2 different values of the vanadium resistivity are compared to the effective BCS DOS used in the text for the theoretical interpretation. Both small a) and large b) Dynes parameter in the V layer are considered.}
		\label{FigS2}
	\end{centering}
\end{figure}
When the superconducting ring is made of a bilayer, the situation is more involved (as we detail in the next subsection): In first approximation it is possible to model the total kinetic inductance of the bilayer as the parallel of the kinetic inductance of the two layers. This simple calculation shows how the inclusion of the Al underlayer provides a suitable geometry for the good phase biasing of the device, independently of the specific properties of the vanadium layer.

%  \begin{figure}[h]
%  	\begin{centering}
% 		\includegraphics[width=0.3\textwidth]{FigS1.pdf}
% 		\caption{Schematics of the V-Al bilayer. }
%		\label{Fig1}
%\end{centering}
% \end{figure}

%\begin{comment}

\section*{Bilayer modeling}
The spectral properties of the V-Al bilayer in the dirty limit can be modeled within the Usadel formalism\cite{Usadel1970}. %In particular the Usadel equation must be solved with the proper boundary condition and the position dependent order parameter must be computed self consistently.
The problem formulation is similar to the one given by Fominov and Feigel'man for the properties of a thin NS bilayer\cite{Fominov2003}. 
In the numerical computation we model the bilayer as a superconducting strip with total thickness $t=100$ nm and we assume a ratio 1:1 ($t_\textrm{Al}=t_\textrm{V}=50 $ nm) between the two layers, accordingly to the experimental  realization (Fig. 1). We assume a clean interface between the two layers. 

A parameter relevant for the properties for the bilayer is 
\begin{equation}
p  =  \frac{t_{\mathrm{V}}\rho_{\mathrm{Al}} D_{\mathrm{Al}}\lambda_\textrm V}{t_{\mathrm{Al}}\rho_{\mathrm{V}} D_V \lambda_\textrm {Al}}
\end{equation}
where $\rho_X$ and $D_X$ are the normal state resistances and the diffusion constants of the materials $X=$Al,V (Einstein relation $D_X^{-1}=\rho_X e^2 \nu_X$ is assumed and $\nu_X$ is the density of states at the Fermi level). The coupling constants in the two superconducting layers $\lambda_X=-\ln(\Delta_X/2E_\textrm c^X)$ depend in the weak coupling limit on the cutoff energy $E_\textrm c^X\sim k_\textrm b\theta_\textrm D^X$ where $\theta_\textrm D^X$  is the Debye temperature. % $=400$ K which is assumed to be the same for both materials. 

%($\theta_\textrm D^\textrm{Al}$=430 K, $\theta_\textrm D^\textrm{V}$=340 K). 
The density of states at the Fermi level $\nu_X=N_X(E_F) d_X/M_X$ are taken from the literature. Here $N_X$ is the density of states at the Fermi level for atom ($N_\textrm{Al}(E_F)=0.208\ \textrm{eV}^{-1} $ , $N_\textrm{V}(E_F)=1.31\ \textrm{eV}^{-1} $)\cite{McMillan1968}, $d_X$ is the mass density ($d_\textrm{Al}=2.7 \ \textrm{g/cm}^3 $ , $d_\textrm{V}=6.0 \ \textrm{g/cm}^3$) \cite{haynes2016crc} and $M_X$ is the atomic mass ($M_\textrm{Al}=26.98 \ \textrm{u} $ , $M_\textrm{V}=50.94 \ \textrm{u}$) \cite{Meija2016}. The Debye temperature is assumed to be the same for both materials $\theta_D^\textrm{Al}=\theta_D^\textrm{V}=400$ K.

For the Al layer we choose a critical temperature equal to the bulk value $T_\textrm C^\textrm{Al}=1.2$ K, corresponding to a zero temperature order parameter $\Delta_\textrm{Al}=178\ \mu$eV, and typical resistivity  $\rho_\textrm{Al}= 5\ \mu\Omega\ \textrm{cm}$ and Dynes parameter $\Gamma_\textrm{Al}/\Delta_\textrm{Al}=10^{-4}$  obtained through electron beam evaporation.

As already stated before, the properties of the vanadium deposited through electron beam evaporation are extremely sensitive to the evaporation conditions. In accordance with the discussion in the previous section, we consider  $\rho_\textrm{V}= 5\ \mu\Omega\ \textrm{cm}$ and $\rho_\textrm{V}= 25\ \mu\Omega\ \textrm{cm}$ as minimal and maximal resistivity in the numerical computation. Similarly apply to the critical temperature of the vanadium, which can be significantly smaller than the bulk value\cite{Noer1975}, depending on the evaporation rate. In our numerical computation we set $T_c^\textrm{V}=3.5$ K, which is reasonable due to the low evaporation rate and previous realizations \cite{Quaranta2011}.  Finally we consider two cases for the Dynes parameter of vanadium: a very ideal situation $\Gamma_\textrm{V}/\Delta_\textrm{V}=10^{-4}$ and an extremely leaking layer $\Gamma_\textrm{V}/\Delta_\textrm{V}=0.6$. The latter seems to describe better the results of our experiment as we show in Fig.2, where the DOS at the bottom of the Al layer is compared to the effective BCS DOS used in the main text. 
In particular the resistivity of the Vanadium plays a role in the determination of the energy gap of the bilayer, but does not affect significantly the subgap density of states. In particular, the large subgap conductance observe in the experiment must be associated to an high effective Dynes parameter in the V layer even in this model. 
Notably, the results compare quite well with the effective BCS model used in the main text.

%\end{comment}
In this model, the kinetic inductance of the bilayer is evaluated as $L_\textrm{kin}^{R}=\hbar/2e I_S' (\phi)$, where the supercurrent dispersion $I_S(\phi)$ is computed starting from the solution of the Usadel equation.
An approximate expression for ultrathin layers\cite{Fominov2003} can be obtained in the Cooper limit\cite{Cooper1961}, where the superconducting energy gap is homogeneous along the bilayer\. The kinetic inductance of the ring is therefore given by the parallel of the kinetic inductances of the two layers:

\begin{equation}
\frac{l}{w L_\textrm{kin}}
=\left(   \frac{d_{\mathrm{Al}}}{\rho_{\mathrm{Al}}} + \frac{d_{\mathrm{V}}}{\rho_{\mathrm{V}}}  \right)
\frac{\pi\Delta_{\mathrm{V}}}{\hbar}\Bigl(\frac{\Delta_{\mathrm{V}}}{\Delta_{\mathrm{Al}}}\Bigr)^{1/(1+p)}
\end{equation}

In Tab. \ref{TabSupp}, we see how the approximate expressions for the kinetic inductance compare to the values obtained through the rigorous calculation. Generally, the approximation underestimates the kinetic inductance somewhat, and does not take nonzero Dynes parameters into account.
\begin{table}[h]
	\caption{Kinetic inductance of the V-Al bilayer. The values computed numerically $L_\textrm{kin}^\textrm{num}$ are compared with the  approximate expression $L_\textrm{kin}^\textrm{approx}$ for the numerical computation parameters chosen above.}
	\begin{tabular}{c c c c}
		\hline
		
		\noalign{\smallskip}
		
		$\Gamma_\textrm V/\Delta_\textrm V$ & $\rho_V(\mu\Omega$ cm) & $L_\textrm{kin}^\textrm{num}$ (pH) & $L_\textrm{kin}^\textrm{approx}$ (pH) \\ 
		\hline
		10$^{-4}$ & 5 &  1.36 & 1.31\\ %0.227  & 0.219 \\ \hline\hline
		10$^{-4}$ & 25 & 2.54 & 2.19\\ %& 0.423   & 0.365  \\ 
		0.6 & 5 &  2.06  & 1.31\\ % &0.343  & 0.219 \\
		0.6 & 25 & 3.67 & 2.19\\%& 0.612 & 0.365\\ 
		\vspace{0.1cm}
		\vspace{0.2cm}
	\end{tabular}
	\label{TabSupp}
\end{table} 

\section*{Impact of the finite width of the probe}
\begin{figure}[h]
	\begin{centering}
		\includegraphics[width=\textwidth]{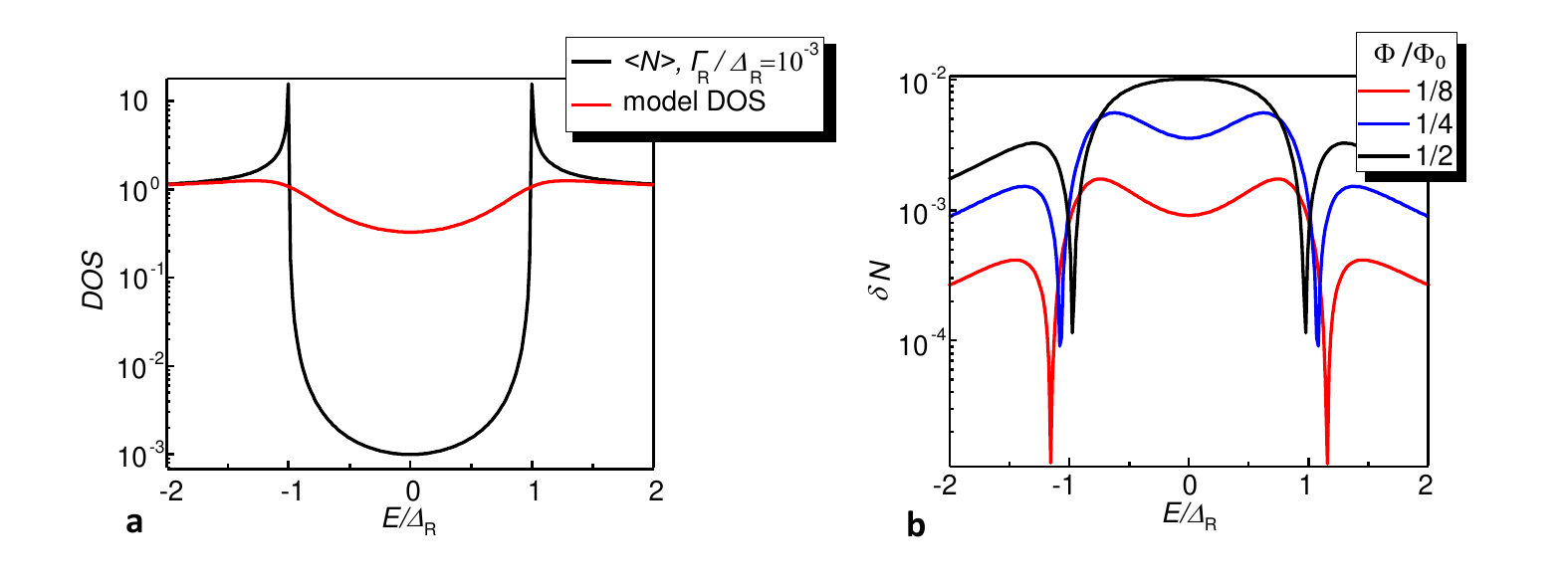}
		\caption{Impact of the finite width of the probe on the theoretical description. 
			(a)The averaged density of states over the finite width of the probe for an ideal ring with small Dynes parameter is compared with the effective model used in the main text at $\Phi=0$. 
			(b)Absolute value of the relative deviation of the zero-width probe approximation for the density of states induced in the nanowire VS quasiparticle energy. Parameters are $\Gamma_\textrm R/\Delta_\textrm R=0.35$, $x_0/L=0.25$ and $w=0.2 L$}
		\label{FigS3}
	\end{centering}
\end{figure}

In the main text is stated that, in order to simplify the calculation, we disregard the finite width of the probe. First, we show that the high subgap conductance observed in the differential conductance curves is not originated by the finite extension of the probe. In Fig. \ref{FigS3}, panel a) we compare the effective DOS used in the main text with the DOS obtained after averaging over the probe width $<N>=\frac{1}{w}
\int_{x_0-w/2}^{x_0+w/2}N(E,\Phi,x)dx$, where we set an ideal Dynes parameter $\Gamma_\textrm R/\Delta_\textrm R=10^{-3}$. The subgap conductance in the latter case is too small to explain the experimental results. Then we quantify the 
relative deviation between the simplified expression $N(E,\Phi,x_0)$ and the integrated expression $<N>$ through the figure of merit 
\begin{equation}
\delta N(E)=\frac{1}{N(E,\Phi,x_0)}\left|N(E,\Phi,x_0)-<N>\right|.
\end{equation} 

In Fig.\ref{FigS3} panel b) we plot this function for different values of $\Phi\neq 0$ (there is no deviation at $\Phi=0$). We see that the maximum relative deviation is always smaller or equal than 1$\%$.

\section*{Theoretical flux dependence}
\begin{figure}[h]
	\begin{centering}
		\includegraphics[width=\textwidth]{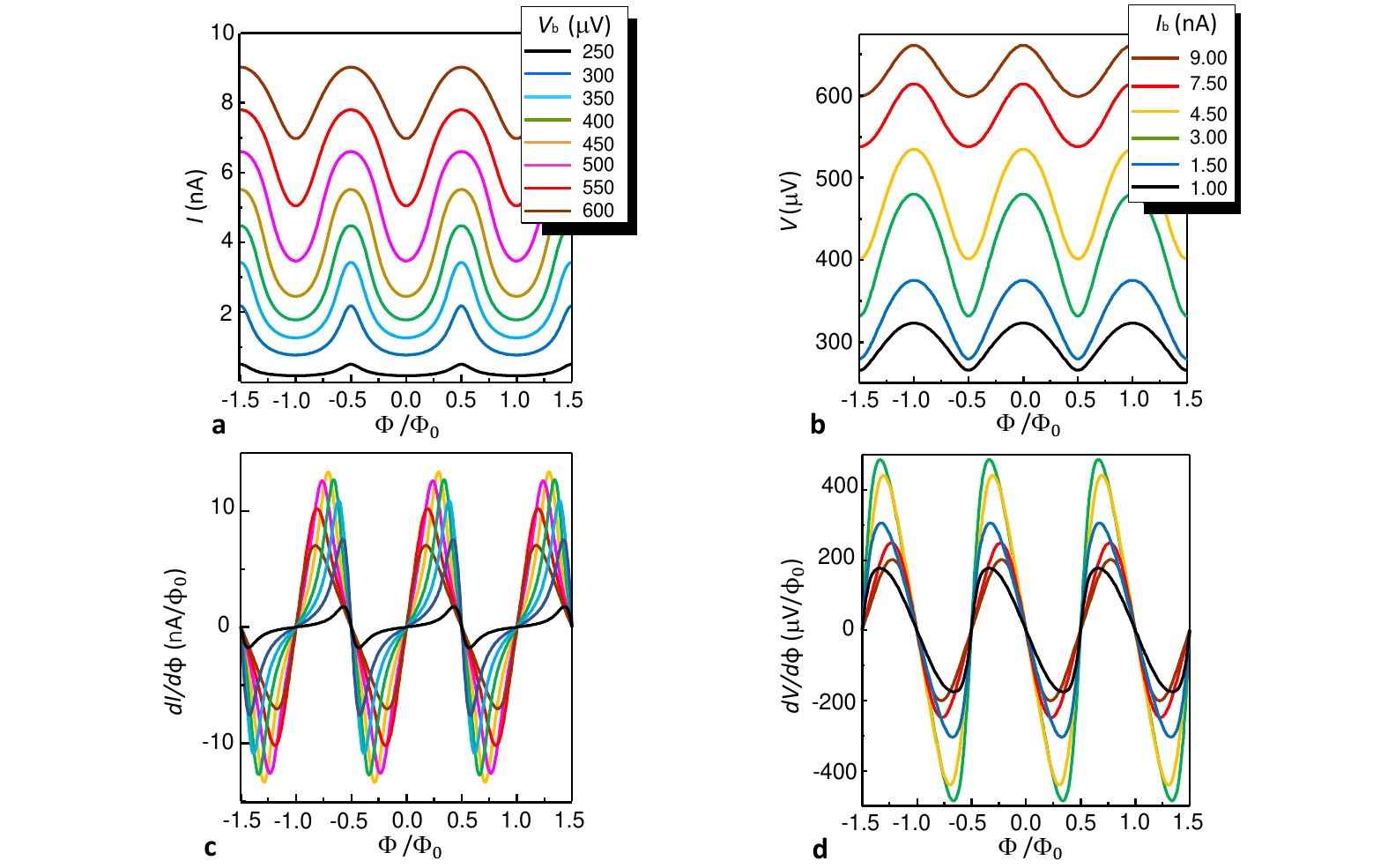}
		\caption{Theoretical interferometric behaviour. (a) Current modulation $I(\Phi )$ for different values of bias voltage $V_\textrm b$ applied to the tunnel junction. (b) Voltage modulation curves $V(\Phi)$ at different values of the biasing current $I_\textrm b$ through the junction. (c) and (d) Flux-to-current $dI/d \Phi $ and flux-to-voltage $dV/d \Phi $ transfer functions, obtained by numerical differentiation of $I( \Phi )$ and $V( \Phi )$, respectively.}
		\label{FigS4}
	\end{centering}
\end{figure}

For completeness, in this section we discuss the theoretical flux dependence obtained from the theoretical model adopted throughout the main text. The plots corresponding to the panel of the Fig. 3 of the main text are displayed in Fig. \ref{FigS4}  We note that the comparison with the experimental data is certainly less satisfactory compared to the differential conductance curves. In particular the predicted oscillation is larger than the observed (especially at larger voltages/currents) and the curves are quite smoother around $0.5\ \Phi_0 + n\Phi_0 $. Notably, despite these deviations, the maximum  current-to-flux and voltage-to-flux transfer functions are close to the one observed in the experiment.
This plot explain why a large deviation in the temperature evolution of the swing is observed in the theoretical curves in Fig. 5 of the main text, whereas the temperature evolution of the maximum transfer function works better. 

%FIGURA1BIS%%%%%%%%%%%%%%%%

%\bibliographystyle{naturemag-doi}
%\bibliography{ForReferees}

\end{document}